# Enhanced stability of antiferromagnetic skyrmion during its motion by anisotropic Dzyaloshinskii-Moriya interaction

Z. P. Huang, Z. Jin, X. M. Zhang, Z. P. Hou, D. Y. Chen, Z. Fan, M. Zeng, X. B. Lu, X. S. Gao, and M. H. Qin [*]

*Institute for Advanced Materials, South China Academy of Advanced Optoelectronics and Guangdong Provincial Key Laboratory of Quantum Engineering and Quantum Materials, South China Normal University, Guangzhou 510006, China*

**[Abstract]** Searching for new methods to enhance stability of antiferromagnetic (AFM) skyrmion during its motion is an important issue for AFM spintronic devices. In this work, we numerically study the spin-polarized current-induced dynamics of a distorted AFM skyrmion based on the Landau-Lifshitz-Gilbert simulations of the model with an anisotropic Dzyaloshinskii-Moriya (DM) interaction. It is demonstrated that the DM interaction anisotropy induces the skyrmion deformation, which suppresses the distortion during the motion and enhances the stability of the skyrmion. Moreover, the effect of the DM interaction anisotropy on the skyrmion velocity is investigated in detail, and the simulated results are further explained by Thiele's theory. This work unveils a promising strategy to enhance the stability and the maximum velocity of AFM skyrmion, benefiting future spintronic applications.

[*]Email: qinmh@scnu.edu.cn

**1 Introduction** Skyrmions are attracting more and more attentions due to their potential applications in future spintronic devices[1-5], especially considering their particular merits including the nanoscale size, the topological protection, and the ultralow critical drive current.[1,6] Specifically, skyrmions are topological defects with vortex-like spin structures which have been experimentally reported in a series of chiral magnets[1,7,8] and heavy metal/ferromagnetic films[9,10]. In these materials, the Dzyaloshinskii-Moriya (DM) interactions[11,12] breaking the inversion symmetry are essential in stabilizing the skyrmion lattice phases. Furthermore, it has been theoretically predicted that skyrmions could exist in frustrated magnets[13,14], and the prediction has been experimentally realized in frustrated kagome $Fe_3Sn_2$ which hosts skyrmionic magnetic bubbles[15].

Moreover, the dynamics of ferromagnetic skyrmions has been extensively investigated, and other external stimuli such as gradient magnetic[16]/electric fields[2,17] and spin waves[18] have been proposed to efficiently drive skyrmions. However, the skyrmion Hall motion[19,20] is induced due to the Magnus forces acting on the skyrmions, which prohibits a precisely control of the motion and goes against future applications. For example, a ferromagnetic skyrmion could be restricted by element edges of related devices, limiting the stable data store and transmission. Interestingly, this problem could be well solved through replacing ferromagnetic skyrmions by antiferromagnetic (AFM) skyrmions which have been theoretically predicted in several AFM systems.[21-23] Concretely, an AFM skyrmion is comprised of two coupled topological spin textures with opposite topological numbers[24-26] as depicted in Fig. 1(a), and the Magnus forces acting on the two sub-lattices are well canceled. As a result, the skyrmion Hall motion is completely suppressed, and AFM skyrmions can move straightly along the driving stimulus direction.

Subsequently, the spin-polarized current-driven dynamics of AFM skyrmions has been clarified.[24-27] Interestingly, the minimum driving current density is about two orders smaller than ferromagnetic skyrmion, and the velocity is about one order larger under a same current density.[25] These important reports definitely demonstrate the great potential of the AFM skyrmions for future racetrack memories, while their stability during the motion deserves to be further enhanced. Concretely, under a high drive current density, the AFM skyrmion is deformed from a circle shape to an ellipse shape during its motion, and even stretched to two domain walls, causing unexpected information loss.[24,25] Thus, Searching for new methods to enhance stability of AFM skyrmion during its motion is an important issue for AFM spintronic applications.

On the other hand, the deformation of the ferromagnetic skyrmion has been experimentally reported in MgO/CoFeB/Pt[28] and strained FeGe[29] films, respectively. It is revealed in our earlier work that the anisotropic DM interaction in strained FeGe plays an essential role in the skyrmion deformation, and results in an anisotropic dynamics of the distorted skyrmion.[30] Most recently, strong DM interaction anisotropy induced by compressive strain was also reported in Co/Pt multilayers.[31] In some extent, DM interaction anisotropy could be also induced by applying uniaxial or anisotropic strain in AFM film, and results in a deformation of AFM skyrmion. More importantly, the distortion could be appropriately modulated to suppress the deformation of the AFM skyrmion during its motion. As a result, such a distorted AFM skyrmion probably has an enhanced stability to stand up to high current and speed. Therefore, the study of the dynamics of distorted AFM skyrmion is essential both in basic physical research and in application potential.

In this work, we study the motion of the distorted AFM skyrmions driven by a spin-polarized current based on the Landau-Lifshitz-Gilbert (LLG) simulations of a two-dimensional model with the anisotropic DM interaction. It is demonstrated that the DM interaction anisotropy induces the skyrmion deformation, which significantly suppresses the distortion and enhances the stability of the skyrmion during the motion. Moreover, the effect of the DM interaction anisotropy on the skyrmion velocity has been investigated in detail, and the simulated results are explained by Thiele's theory.

## 2 Computational details

We study the classical AFM model with the anisotropic DM interaction on the two-dimensional square lattice:

$$H = J\sum_{\langle i,j \rangle}\mathbf{m}_i \cdot \mathbf{m}_j - \sum_i \left( D_x \mathbf{m}_i \times \mathbf{m}_{i+x} \cdot \hat{y} - D_y \mathbf{m}_i \times \mathbf{m}_{i+y} \cdot \hat{x} \right) - K\sum_i \left( \mathbf{m}_i^z \right)^2, \quad (1)$$

where $\mathbf{m}_i$ is the unit vector of the magnetic moment at site $i$, $\mu_i = -\hbar\gamma S_i$[32] with $S_i$ being the atomic spin, $\gamma$ the gyromagnetic ratio, $\hbar$ the reduced Plank constant. The first term is the AFM exchange interaction between the nearest neighbors with $J = 1$, the second term represents the anisotropic interfacial DM interaction with the interaction anisotropy defined by $\eta = D_y/D_x - 1$ (depicted in the supporting information S1), and the last term is the perpendicular magnetic anisotropy with the anisotropic constant $K = 0.25J$. Here, the interfacial DM interaction which stabilizes the Néel-type skyrmion is considered, and the bulk DM interaction stabilizing the Bloch-type skyrmion exhibits similar results.

The dynamics induced by a spin-polarized current in the current-perpendicular-to-plane (CPP) geometry is investigated by solving the updated LLG equation,

$$\frac{d\mathbf{m}_i}{dt} = -\gamma \mathbf{m}_i \times \mathbf{H}_i + \alpha \mathbf{m}_i \times \frac{d\mathbf{m}_i}{dt} + \frac{\gamma}{a} u \mathbf{m}_i \times (\mathbf{p} \times \mathbf{m}_i), \tag{2}$$

where $\mathbf{H}_i = -(1/\mu_i)\partial H/\partial \mathbf{m}_i$ is the effective field, $\alpha$ is the Gilbert damping coefficient, $u$ is the spin transfer torque coefficient, $\mathbf{p}$ represents the electron polarization direction, and $a$ is the lattice constant. Here, $u = \hbar j P/2eM_s$ with $j$ the current density, $P$ the spin polarization rate, $e$ the elementary charge, and $M_s = \hbar\gamma S/a^3$ the saturation magnetization[32]. Without loss of generality, we set $\hbar = \gamma = S = a = 1$, and the time $t$, velocity $v$ and current density $j$ can be converted into SI units through $t = \hbar S/J$, $v = Ja/\hbar S$ and $j = Je/\hbar a^2$, respectively.

The initial spin configurations are obtained using the Monte Carlo simulations performed on an $24 \times 24$ square lattice with the periodic boundary condition, and are sufficiently relaxed by solving the LLG equation using the fourth-order Runge-Kutta method. Subsequently, the spin dynamics driven by the spin-polarized current are investigated, and the simulated results are further confirmed and explained using the approach proposed by Thiele[33]. The displacement of the AFM skyrmion is characterized by the position of its center $(R_x, R_y)$:

$$R_\mu = \frac{\int dxdy \left[\mu \cdot (1-m^z)\right]}{\int dxdy (1-m^z)}, \quad \mu = x, y. \tag{3}$$

Then, the velocity is numerically calculated by $(v_x, v_y) = (dR_x/dt, dR_y/dt)$.

### 3 Results and discussion

**3.1 Static spin configurations of isolated AFM skyrmions** Fig. 1(a) gives the static spin configuration of a single AFM skyrmion in the absence of the DM interaction anisotropy $\eta = 0$ for $D_x = 0.4$. The AFM skyrmion clearly exhibits arbitrary rotation symmetry and can be decoupled into two isolated ferromagnetic skyrmions with opposite topological numbers, as clearly shown in the bottom of Fig. 1(a) (Generally, 1 for the left ferromagnetic skyrmion and -1 for the right skyrmion). In order to help one to understand the configuration more clearly, the *z*-components of the two-sublattice magnetic moments along the central *x*-axis and *y*-axis are presented in Fig. 2(a) and 2(b), respectively. It is shown that the AFM skyrmion is axisymmetric and with a size ~8 lattices.

When a lattice distortion is generated by applied strain, the DM interaction anisotropy could be induced and efficiently modulates the AFM skyrmion structure, as shown in Fig. 1(b) where gives the spin snapshot for $\eta = 0.15$ and $D_x = 0.4$. It is clearly shown that the skyrmion size is significantly enlarged due to the enhanced DM interaction $D_y$. More importantly, the AFM skyrmion is obviously deformed from the circle shape at $\eta = 0$ to the elliptical shape at $\eta = 0.15$ with the long axis along the *x*-direction with weak DM interaction. Moreover, the deformed AFM skyrmion can be also decoupled into two distorted ferromagnetic skyrmions whose topological charges are not changed, as shown in the bottom of Fig. 1(b). Fig. 2(c) and 2(d) give respectively the two-sublattice *z*-components of the spins along the central *x*-axis and *y*-axis, which show the increase of the skyrmion size and a deformation up to ~25% of the skyrmion. Moreover, near the center of the AFM skyrmion, the nearest neighboring spins arrange antiparallel with each other.

**3.2 Enhanced stability of AFM skyrmion during the motion** Subsequently, the motion of the AFM skyrmion driven by the spin current in the CPP configuration is studied in detail. When the current is applied, the skyrmion moves straightly along the *x*-axis direction without any skyrmion Hall motion. However, in the absence of the DM interaction anisotropy, the AFM skyrmion is quickly deformed to an elliptical shape during its motion with the long axis along the *y*-direction, as shown in Fig. 3(a) where gives the spin configuration for $\eta = 0$ under the current density $j = 0.2$. Moreover, as *j* increases above 0.4, the skyrmion is not stable and stretched to two AFM domain walls, as depicted in Fig. 3(b).

Interestingly, the destabilization of the AFM skyrmion during the motion can be significantly suppressed by a weak DM interaction anisotropy. For example, the skyrmion configuration is rather stable under $j = 0.4$ for $\eta = 0.025$, although a large deformation of the skyrmion still occurs, as shown in Fig. 3(c). Thus, it is clearly indicated that the skyrmion stability during the motion can be enhanced by the strain-induced DM interaction anisotropy. Furthermore, other values of $\eta$ on the stability are also investigated, and the simulated results are summarized in Fig. 3(d). The critical destabilize current $j_c$, beyond which the AFM skyrmion is not stable any more during the motion, is significantly increased with $\eta$. For example, $j_c$ increases from 0.38 at $\eta = 0$ to ~0.48 at $\eta = 0.2$, generating a 25% uplift.

In order to confirm and well explain our simulations, a comparison between the simulations and analytical calculations is indispensable, noting that the skyrmion velocity could be estimated based on the Thiele's theory[33-35]. For brevity, only the derived velocity is given here, whereas the detailed derivation is provided in the Supporting information S2. Based on the Thiele's theory, the velocity is estimated by

$$v = \frac{\gamma I_{xy}}{a\alpha \Gamma_{xx}} u, \tag{4}$$

where $I_{xy}$ is the component of the driving force tensor[36] given by

$$I_{\mu\nu} = \frac{1}{4\pi} \int \left(\frac{\partial n}{\partial \mu} \times n\right)_\nu dxdy, \tag{5}$$

with the Néel vector $\mathbf{n}$[37-41], $\Gamma_{xx}$ is the component of the dissipative tensor given by

$$\Gamma_{\mu\nu} = \frac{1}{4\pi} \int \frac{\partial n}{\partial \mu} \cdot \frac{\partial n}{\partial \nu} dxdy. \tag{6}$$

The analytically calculated and the LLG simulated velocities as functions of $j$ for $\eta = 0$ are presented in Fig. 4(a). With the increase of $j$, the spin transfer torque is enhanced, which drives the skyrmion to move fast. The analytical and simulated results are in well consistent with each other under weak $j < 0.2$, while slightly deviate from each other under high $j$. It is noted that Eq. (4) is derived based on the assumption that the spin configuration is not changed during the motion and $\Gamma_{xx}$ always equals to $\Gamma_{yy}$, while the deformation of the skyrmion under high $j$ is completely ignored. As a matter of fact, for a deformed skyrmion, the difference between $\Gamma_{xx}$ and $\Gamma_{yy}$ could be very large, as clearly shown in Fig. 4(b) where presents the simulated $\Gamma_{xx}$ and $\Gamma_{yy}$ as functions of $j$. With the increase of $j$, $\Gamma_{xx}$ is significantly increased attributing to the skyrmion deformation, while $\Gamma_{yy}$ is almost unchanged. Thus, the analytical calculation is performed on a skyrmion whose size much larger than the actual skyrmion. As a result, the analytically calculated velocity is larger than the simulated velocity under high $j$, because a large skyrmion is generally with a high mobility. However, the perfect consistence between the simulations and calculations under weak $j$ clearly demonstrates the reliability of our simulated results.

**3.3 Impact of the DM interaction anisotropy on the velocity** Undoubtedly, the dependence of the skyrmion velocity $v$ on $\eta$ is very important for future applications. Fig. 5(a) shows the skyrmion velocity depending on $\eta$ under $j = 0.1$ and $j = 0.4$, which exhibits two different behaviors for the cases of small and large $j$. On one hand, under small $j = 0.1$, the velocity first increases with $\eta$ to a maximum value at $\eta \sim 0.13$, and then decreases as $\eta$ further increases. This phenomenon could be understood from the following aspects. For a fixed $j$, the velocity is mainly determined by $I_{xy}/\Gamma_{xx}$, as revealed in Eq. (4). As $\eta$ increases, the skyrmion size is enlarged, and $I_{xy}/\Gamma_{xx}$ is increased as shown in Fig. 5(b), resulting in the increase of $v$. Moreover, the anisotropy-induced deformation of the skyrmion also contributes to the motion. Under small $j$, the deformation plays an important role for large $\eta$, which suppresses the driving force and reduces the speed of the skyrmion, as revealed in our simulations. This phenomenon is also available for other parameter values, as shown in Fig. 5(c) where presents the simulated $v$-$\eta$ curves for various $K$ under $j = 0.1$. For a fixed $\eta$, $v$ decreases with the increase of $K$ due to the reduced skyrmion size. Moreover, the effect of $\eta$ on $v$ is also suppressed by the enhanced $K$, resulting in the fact that the critical $\eta$ is increased while the maximum velocity is decreased, while the deformation of the skyrmion hardly be changed.

On the other hand, under large $j = 0.4$, the velocity is slightly increased as $\eta$ increases. In this case, the skyrmion is significantly deformed during its motion even for large $\eta$, and the long axis of the skyrmion changes from the *x*-direction to the *y*-direction. Thus, the enlargement of the skyrmion size with $\eta$ mainly contributes to the slight increases of $v$ and $I_{xy}/\Gamma_{xx}$. More importantly, the DM interaction anisotropy can be used to suppress the skyrmion deformation during the motion, extensively enhancing the stability of the skyrmion.

Fig. 5(d) shows the simulated (empty points) and calculated (solid lines) $v$ as functions of $j$ for various $\eta$, which are well consistent with each other. The skyrmion moves fast for large $\eta$ under a fixed $j < 0.15$, while hardly be affected by $\eta$ under large $j$. The results could be also explained by the Thiele's theory, as shown in Fig. 5(e) where gives the simulated $I_{xy}/\Gamma_{xx}$. Under small $j$, $I_{xy}/\Gamma_{xx}$ is enlarged with the increase of $\eta$, speeding up the skyrmion. The $I_{xy}/\Gamma_{xx}(j)$ curves are gradually merged with the increase of $j$, and the velocity less depends on $\eta$ under large $j$.

Subsequently, we intend to discuss the results in the practical units. For the parameter set $(J, a)$ = (1meV, 0.4nm) in the absence of the DM interaction anisotropy, the critical current density is estimated to be $j_c$ ~ 2.85 × $10^{12}$A/m$^2$, well consistent with the earlier report where $j_c$ ~ 3.0 ×$10^{12}$A/m$^2$ is obtained. Interestingly, this work demonstrates that a weak $\eta = 0.15$ could enhance the skyrmion stability during the motion and enlarge $j_c$ by nearly 20%, which speeds up the AFM skyrmion by 7%. As a matter of fact, the DM interaction anisotropy and skyrmion deformation could be induced through applying uniaxial strain or anisotropic strain, which has been experimentally reported in FeGe.[29] It is demonstrated that even a small anisotropy strain ~0.3% could induce a large skyrmion deformation ~20%. Furthermore, the anisotropic DM interaction has been reported recently in ultra-thin epitaxial Au/Co/W, and isolated elliptical skyrmions are expected.[42] Thus, deformed AFM skyrmions may play an important role for future spintronic devices because of their stability during the motion.

**3.4 Skyrmion motion driven by in-plane current** So far, the spin current in the CPP geometry driven AFM skyrmion motion with the DM interaction anisotropy has been clarified, and we pay attention to the case of the spin current in the Current-in-plane (CIP) geometry. In this case, the LLG equation is updated to[39,43]

$$\frac{d\mathbf{m}_i}{dt} = -\gamma\mathbf{m}_i \times \mathbf{H}_i + \alpha\mathbf{m}_i \times \frac{d\mathbf{m}_i}{dt} + \gamma u\mathbf{m}_i \times \left(\frac{\partial \mathbf{m}_i}{dx} \times \mathbf{m}_i\right) - \beta u\left(m_i \times \frac{\partial \mathbf{m}_i}{dx}\right), \quad (7)$$

where the third term in the right side is the adiabatic spin-transfer-torque term, and the last $\beta$ term is the non-adiabatic term. Fig. 6(a) presents the simulated velocity (empty points) as a function of $j$ for various $\beta$ for $\eta = 0$, clearly demonstrates the increase of $v$ with $j$ and/or $\beta$, well consistent with the earlier analytical theory (solid lines) $v = \beta u/\alpha$.[24] Moreover, the size of the AFM skyrmion is also enlarged under large $j$, resulting in the slight deviation between the simulations and calculations. The simulated velocity driven by the spin current in the CPP geometry is also presented, which is significantly larger than the CIP driven case under a fixed $j$. Thus, it is demonstrated that the out-of-plane current is more efficient than the in-plane current in driving the AFM skyrmions, similar to the dynamics of ferromagnetic skyrmions[36,43].

At last, we investigate the effect of $\eta$ on the skyrmion speed, and give the corresponding results in Fig. 6(b). Different from the case of CPP geometry, $v$ slightly decreases with the increase of $\eta$ under a fixed $j$. It is noted that the AFM skyrmion is deformed by the introduced $\eta$ with the long axis along the $x$-direction, which is hardly changed due to the comparatively low velocity under the in-plane current. Thus, the $\eta$-induced deformation mainly suppresses the in-plane current driven skyrmion motion. As a matter of fact, similar phenomenon has been observed in the motion of the distorted ferromagnetic skyrmion driven by the in-plane current.[30] Furthermore, the anisotropic dynamical responses of the ferromagnetic skyrmion has been reported, and similar behavior could be also existed in AFM system, which deserves to be further checked.

**4 Conclusion** In summary, we have studied the dynamics of the distorted AFM skyrmion driven by the spin-polarized currents based on the LLG simulations of the model with the anisotropic DM interaction. It is demonstrated that the stability of the skyrmion during the motion can be extensively enhanced by the DM interaction anisotropy, and the critical current $j_c$ is increased by ~20% for the anisotropy magnitude ~0.15. Moreover, the effect of the DM interaction anisotropy on the skyrmion velocity has been investigated, and the simulated results are explained by the Thiele's theory. Thus, this work unveils a promising strategy to enhance the stability of AFM skyrmion during the motion, benefiting future AFM spintronic applications.

**Acknowledgements**:

We sincerely appreciate the insightful discussions with Xichao Zhang and Laichuan Shen from CUHK-Shenzhen, and Jun Chen from Southeast University. The work is supported by the Natural Science Foundation of China (No. 51971096), and the Science and Technology Planning Project of Guangzhou in China (Grant No. 201904010019), and the Natural Science Foundation of Guangdong Province (Grant No. 2019A1515011028).

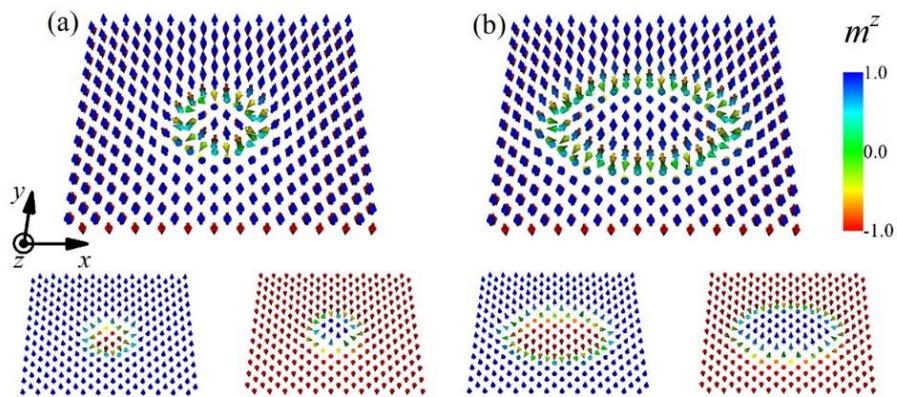

Figure.1. Spin configurations (top) and the two sublattice spin structures (bottom) in the (a) axisymmetric AFM skyrmion for $\eta = 0$, and (b) distorted AFM skyrmion for $\eta = 0.15$.

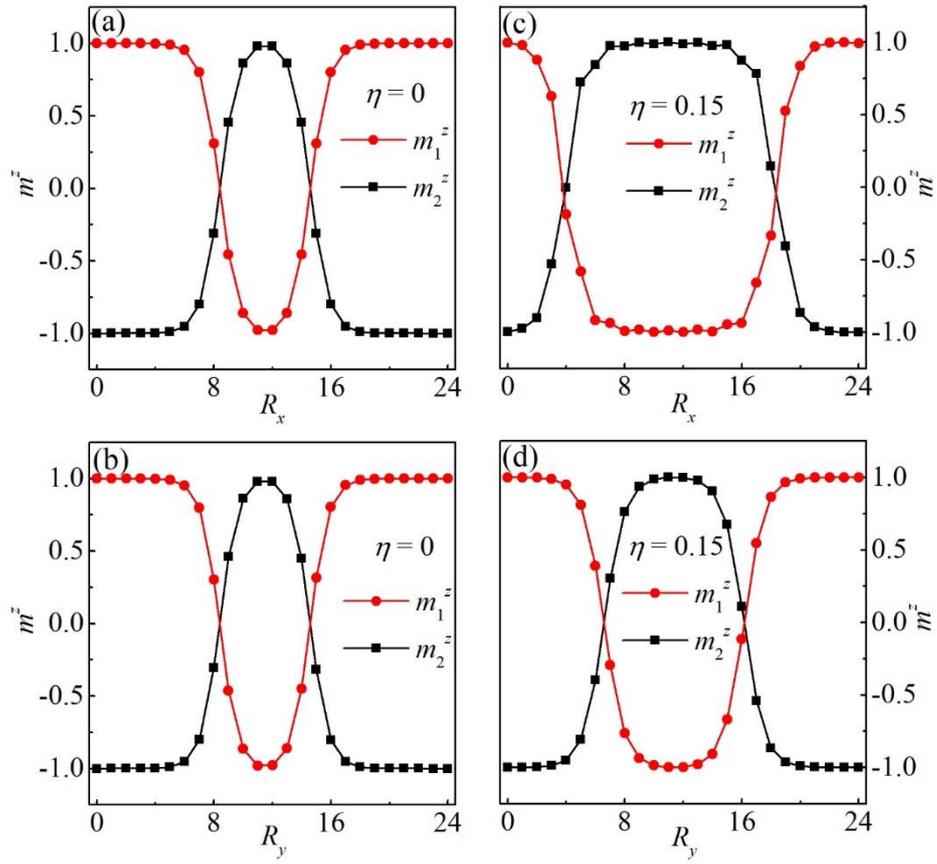

Figure.2. The $z$ components of two-sublattice magnetic moments for $\eta = 0$ ((a) and (b)), and for $\eta = 0.15$ ((c) and (d)) along the central $x$-axis ((a) and (c)) and $y$-axis ((b) and (d)).

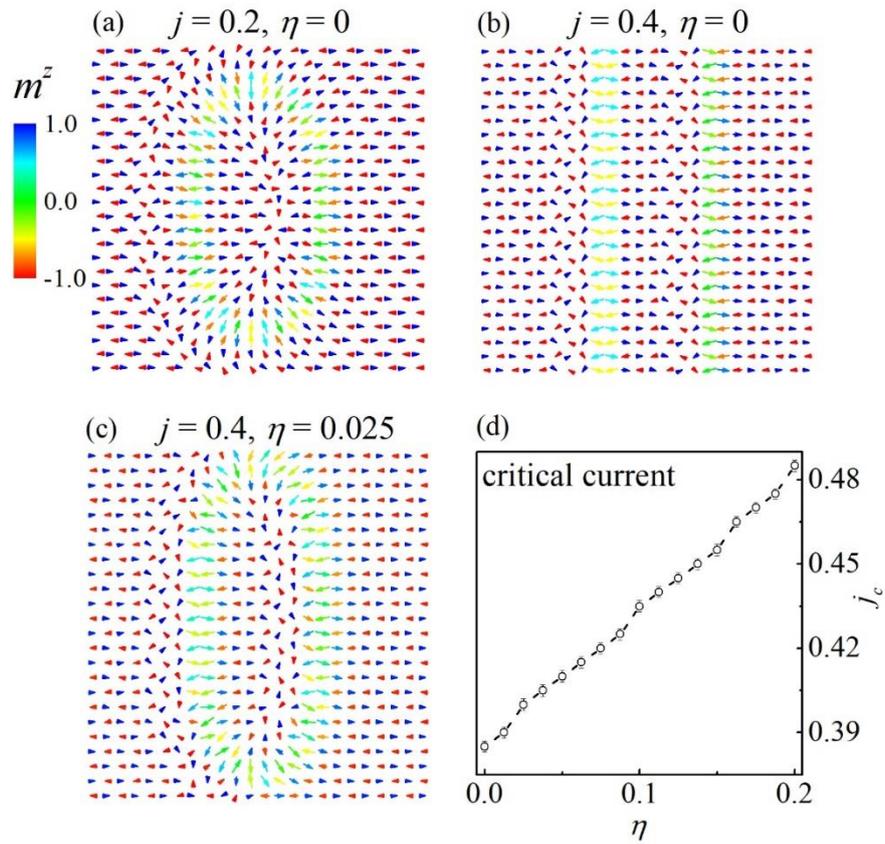

Figure.3. Spin configurations in the (a) distorted AFM skyrmion for $\eta = 0$ under $j = 0.2$, and (b) two domain walls for $\eta = 0$ under $j = 0.4$, and (c) distorted AFM skyrmion for $\eta = 0.025$ under $j = 0.4$. (d) The critical current density $j_c$ as a function of $\eta$.

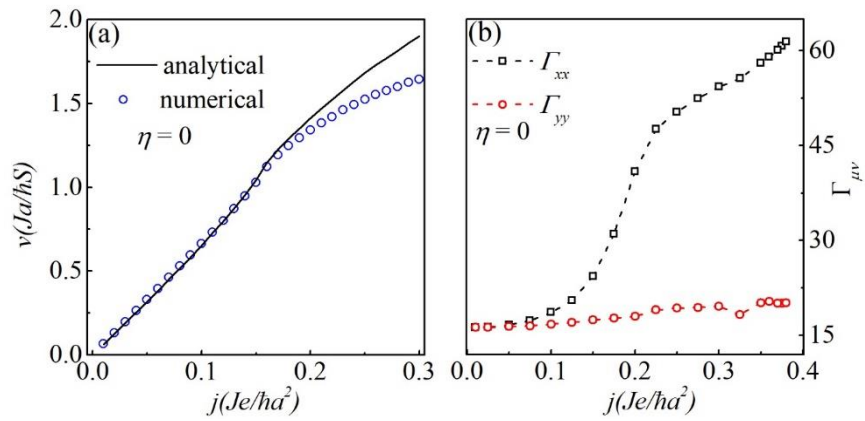

Figure.4. (a) The simulated (empty circles) and analytically calculated (solid line) velocities as functions of $j$, and (b) the components of the dissipative tensor $\Gamma_{xx}$ and $\Gamma_{yy}$ as functions of $j$ for $\eta = 0$.

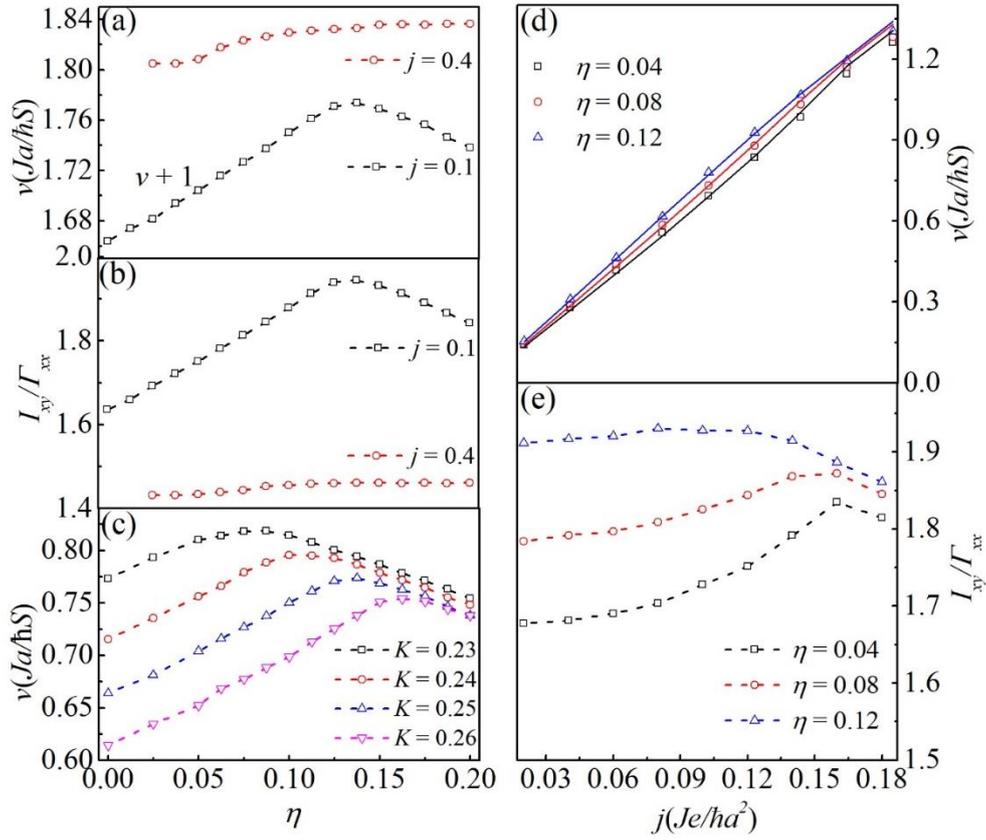

Figure.5. The simulated velocity (a) and $I_{xy}/\Gamma_{xx}$ (b) as a function of $\eta$ under $j = 0.1$ and $j = 0.4$, and the simulated velocity (c) as a function of $\eta$ for various $K$ under $j = 0.1$, and the simulated $v$ (empty points) (d) and $I_{xy}/\Gamma_{xx}$ (e) as a function of $j$ for various $\eta$. The analytically calculated $v$ are also presented in (d) with solid lines.

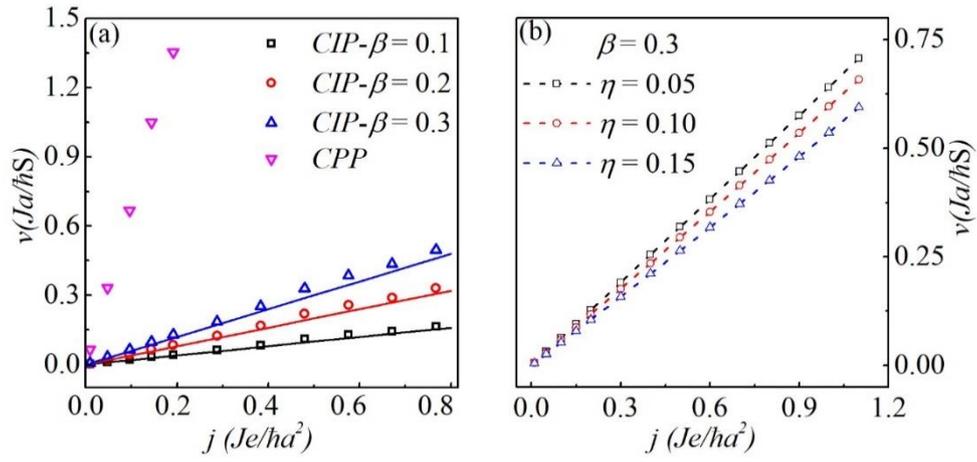

Figure.6. The simulated (empty points) $v$ as a function of $j$ in the CIP geometry for (a) various $\beta$ for $\eta = 0$, and (b) various $\eta$ for $\beta = 0.3$. The analytically calculated $v$ (solid lines) and the velocity driven by spin-current in the CPP geometry are also presented in (a).

# Supporting Information:
# Enhanced stability of antiferromagnetic skyrmion during its motion by anisotropic Dzyaloshinskii-Moriya interaction

Zongpeng Huang, Zhejunyu Jin, Xiaomiao Zhang, Zhipeng Hou, Deyang Chen, Zhen Fan, Min Zeng, Xubing Lu, Xingsen Gao, and Minghui Qin [*]

*Institute for Advanced Materials, South China Academy of Advanced Optoelectronics and Guangdong Provincial Key Laboratory of Quantum Engineering and Quantum Materials, South China Normal University, Guangzhou 510006, China*

## S1. The implementation of anisotropic DM interaction in the model.

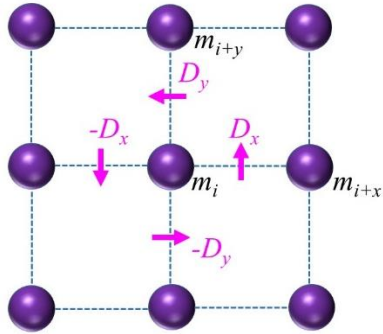

SFig.1. Depiction of the anisotropic DM interaction.

In the absence of the DM interaction anisotropy, $D_x = D_y = D$ is considered with the DM interaction constant $D$. $D_x/D_y$ is the DM interaction coupling between the nearest-neighbor spins along the $x/y$ axis, as clearly depicted in SFig. 1. The DM interaction anisotropy is implemented by introducing a nonzero $\eta = D_y/D_x - 1$.

## S2. The derivation of AFM skyrmion velocity by Thiele's theory

---

[*] Email: qinmh@scnu.edu.cn

For the continuum AFM model, the total magnetization $\mathbf{m}(r, t) = (\mathbf{m}_1(r, t) + \mathbf{m}_2(r, t))/2$ and Néel vector $\mathbf{n}(r, t) = (\mathbf{m}_1(r, t) - \mathbf{m}_2(r, t))/2$ are defined. With the constraints $|\mathbf{n}| = 1$ and $\mathbf{m} \cdot \mathbf{n} = 0$, the model Hamiltonian is updated to

$$H = \int dV \left\{ A_h \mathbf{m}^2 + A(\nabla \mathbf{n})^2 - \left[ D'_x n_z \nabla \cdot \mathbf{n} - D'_y (\nabla \cdot \mathbf{n}) n_z \right] - K' n_z^2 \right\}, \tag{1}$$

where $A_h = 4JS^2/a$ and $A = 2aJS$ are the homogeneous and inhomogeneous exchange constants with $S$ the spin length, respectively. The third term represents the anisotropic interfacial DM interaction constant with $D'_x = D_x S^2$ and $D'_y = D_y S^2$. The last term is the perpendicular magnetic anisotropy with the anisotropic constant $K' = 2KS^2/a$. Subsequently, the magnetic dynamics is described by the following two coupled equations[1,2]:

$$\dot{\mathbf{n}} = (\gamma \mathbf{f}_m - \alpha \dot{\mathbf{m}}) \times \mathbf{n} + \gamma \frac{u}{a} \mathbf{n} \times (\mathbf{m} \times \mathbf{p}), \tag{2a}$$

$$\dot{\mathbf{m}} = (\gamma \mathbf{f}_n - \alpha \dot{\mathbf{n}}) \times \mathbf{n} + (\gamma \mathbf{f}_m - \alpha \dot{\mathbf{m}}) \times \mathbf{m} + \gamma \frac{u}{a} \mathbf{n} \times (\mathbf{n} \times \mathbf{p}), \tag{2b}$$

where $\mathbf{f}_n = -\delta H/\delta \mathbf{n}$ and $\mathbf{f}_m = -\delta H/\delta \mathbf{m}$ are the effective fields.

Following the earlier work[2], we substitute $\mathbf{f}_m$ into Eq. (2a) and combine it with Eq. (2b), and obtain the total magnetization $\mathbf{m}$

$$\mathbf{m} = \frac{1}{2A_h} \left( \frac{1+\alpha^2}{\gamma} \dot{\mathbf{n}} - \alpha \mathbf{f}_n \right) \times \mathbf{n}. \tag{3}$$

Then, this equation is differentiated and combined with Eq. (2b). One gets the equation for the Néel vector $\mathbf{n}$,

$$\frac{1+\alpha^2}{2\gamma^2 A_h} \ddot{\mathbf{n}} = \mathbf{f}_n - \frac{\alpha}{\gamma} \dot{\mathbf{n}} + \frac{u}{a} (\mathbf{n} \times \mathbf{p}), \tag{4}$$

where the dissipative term is small and safely ignored. Taking the scalar product of Eq. (4) and integrating over the space, one gets the Thiele equation[3]

$$\mathbf{a} \cdot \mathbf{M}_{\text{eff}} = \partial_\mu \mathbf{n} \cdot \mathbf{f}_n - \frac{\alpha}{\gamma} \Gamma \mathbf{v} + 4\pi \mathbf{B} \cdot \mathbf{j}_p, \tag{5}$$

where $\mathbf{a}$ is the acceleration, $\mathbf{M}_{\text{eff}} = \frac{(1+\alpha^2)}{2\gamma^2 A_h} \Gamma$ is the effective mass with dissipative tensor $\Gamma$, and $\mathbf{B}$ is driving force tensor related to the current-induced torque. Considering $\mathbf{a} = 0$, the AFM skyrmion steadily moves, and the velocity is estimated by

$$v = \frac{\gamma I_{xy}}{a\alpha \Gamma_{xx}} u \ . \tag{6}$$